\documentclass[a4paper,11pt]{article}
\pdfoutput=1
\usepackage{jinstpub-acc} 

\title{\boldmath A compact, high resolution tracker for cosmic ray muon scattering tomography using semiconductor sensors}

\author[a,1]{F. Keizer,\note{Corresponding author.}}
\author[a]{A. Gorbatch,}
\author[a]{M. A. Parker,}
\author[b]{C. Steer,}
\author[a]{S. A. Wotton}

\affiliation[a]{University of Cambridge, Cavendish Laboratory, JJ Thomson Avenue, Cambridge CB3 0HE, UK}
\affiliation[b]{St Mary's University, Waldegrave Rd, Twickenham TW1 4SX, UK}

\emailAdd{keizer@hep.phy.cam.ac.uk}

\abstract{\\A semiconductor tracker for muon scattering tomography is presented. The tracker contains silicon strip sensors with an $80\,\mu$m pitch, precision mechanics and integrated cooling. The electronic readout of the sensors is performed by a scalable, inexpensive, flexible, FPGA-based system, which is demonstrated to achieve an event rate of $30\,$kHz. The tracker performance  is compared with a Geant4 simulation. A scattering angle resolution compatible with $1.5\,$mrad at the $4\,$GeV average cosmic ray muon energy is demonstrated. Images of plastic, iron and lead samples are obtained using an Angle Statistics Reconstruction algorithm. The images demonstrate good contrast between low and high atomic number materials.
}


\citing{\\F. Keizer \textit{et al} 2018 \textit{JINST} 13 P10028}

\begin{document}
\maketitle
\flushbottom

\clearpage

\section{Introduction}
\label{sec:introduction}

Cosmic ray Muon Scattering Tomography (MST) exploits highly penetrating muons as a probe to passively image high-density, high atomic number objects. Muons travelling through matter undergo multiple Coulomb scattering, where the overall deflection depends on the density, atomic number, path length and muon momentum. Measurements of the angle of deflection can be used to derive an image that distinguishes materials with low, medium and high atomic number \cite{schultz}. The use of muons produced in naturally occurring cosmic ray interactions in the upper atmosphere allows MST to be performed anywhere without the need for on-site particle accelerators. However, the relatively low muon flux of approximately $1\,$cm$^{-2}\,$min$^{-1}$ at sea level means that the image recording time is relatively long. A decrease in recording time is generally a tradeoff with image resolution and contrast.\par

The scattering angle distribution is described by Moliere's multiple scattering theory, where the central $98\,$\% is approximately Gaussian \cite{pdgmst04} with width:
\begin{equation} 
\theta_{0} = \frac{z\cdot13.6\,MeV}{p\, \beta \,c} \,\sqrt{\frac{x}{X_{0}}}\,\Big[1+0.038\cdot\ln{\Big(\frac{x}{X_{0}}\Big)}\Big]
\label{eq:scattering}
\end{equation}
Here, $p$ and $\beta c$ are the momentum and velocity of the muon, $z=1$ is the charge number of the muon and $x$ and $X_{0}$ are the thickness and radiation length of the object respectively. The radiation length is the mean distance over which an electron loses all but $1/e$ of its energy and can be written in approximate form using Dahl\textquoteright s expression \cite{pdgmst04}:
\begin{equation} 
X_{0} = \bigg[\frac{A\cdot 716.4 \,g\,{cm}^{-2}}{Z\, (Z+1)\, \ln{(287/\sqrt{Z})}} \bigg] \,\bigg[\frac{1}{\rho}\bigg]
\label{eq:radiationlength}
\end{equation}
Here, $\rho$, $A$ and $Z$ are the density, atomic weight and atomic number of the material. The approximate $\rho\cdot Z^{2}$ dependence for large $Z$ characterises the sensitivity to dense objects with a high atomic number. The scattering angle distribution in experiment has a contribution from the tracker resolution, as well as the scattering angle $\theta_0$ from the imaged object. The high resolution semiconductor tracker is therefore expected to have increased object material and feature discrimination, which can lead to a decrease in image recording time. \par

MST is of interest for homeland security, such as the scanning of cargo containers for nuclear threats \cite{drift}. Results have been published for MST systems based on scintillator, drift chamber or resistive plate chamber detectors \cite{cript, drift, rpc}. The angular resolution of such systems is determined by the granularity, internal muon scattering and separation of the detector planes. This paper describes the implementation of high resolution silicon technology from particle physics experiments at CERN in the field of MST. The presented tracker uses silicon strip sensors from the ATLAS SemiConductor Tracker (SCT) \cite{barrel}. The $80\,\mu$m silicon strip pitch allows much smaller separation between the inner and outer detector planes while still achieving a similar angular resolution to other published detectors. The tracker therefore opens up the field of compact and portable MST systems that can image objects that are difficult to reach or dangerous to move. The low power, compact and scalable electronic readout system presented in this paper enhances the portability of the tracker. \par

Section \ref{sec:detector} introduces the apparatus and Section \ref{sec:fitting} describes the track fitting and calibration procedures and the Geant4 simulation of the tracker. The imaging algorithm and results are presented in Section \ref{sec:results}. \par

\clearpage

\section{Apparatus}
\label{sec:detector}

The MST system has incoming and outgoing muon tracking stations with multiple silicon layers, arranged as shown in Figure \ref{fig:overview} and Section \ref{sec:tracker}. In this work, the performance of the ATLAS SCT modules, described in Section \ref{sec:sct}, is exploited, which avoids the need for the design, construction and testing of silicon sensors and analog readout electronics. However, the more open MST geometry and lower radiation environment allow simplification of the electronic readout system. The readout used in ATLAS is therefore replaced by the adapter card described in Section \ref{sec:adaptercard} and the readout board in Section \ref{sec:digitalboard}. The readout trigger is provided by the coincidence of scintillator paddle detectors positioned above and below the tracker. Section \ref{sec:trigger} gives an overview of the trigger and readout architecture. The noise characteristic of the system is measured in Section \ref{sec:threshold}.

\subsection{ATLAS SemiConductor Tracker module}
\label{sec:sct}

\begin{figure} [b!]
\centering
\begin{minipage}[t!]{.35\textwidth}
  \centering
  \includegraphics[width=1.\linewidth]{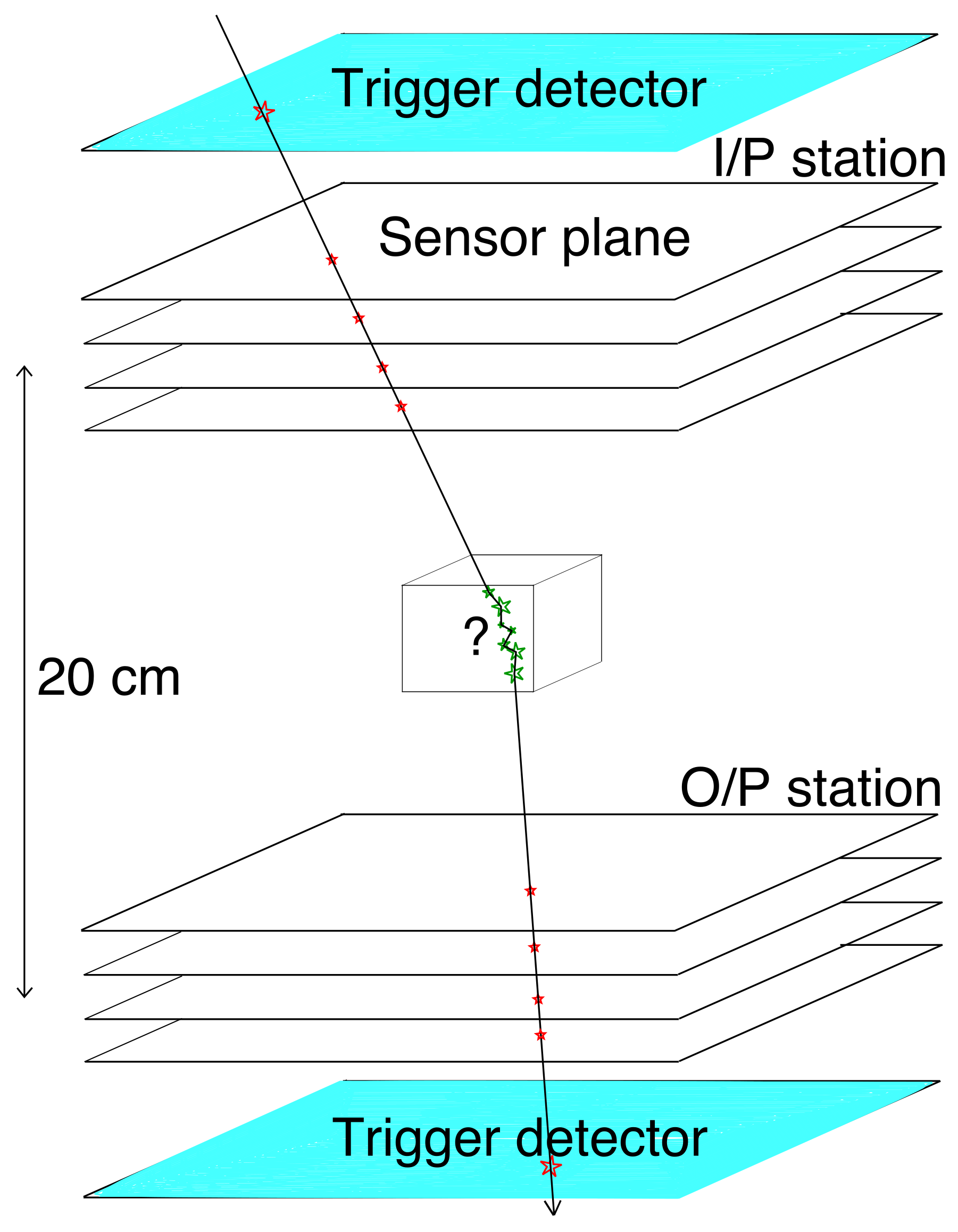}
  \caption{The MST system contains two tracking stations positioned around a sample.}
  \label{fig:overview}
\end{minipage}
\begin{minipage}{.05\textwidth}
\quad
\end{minipage}
\begin{minipage}[t!]{.55\textwidth}
  \centering
  \includegraphics[width=1.\linewidth]{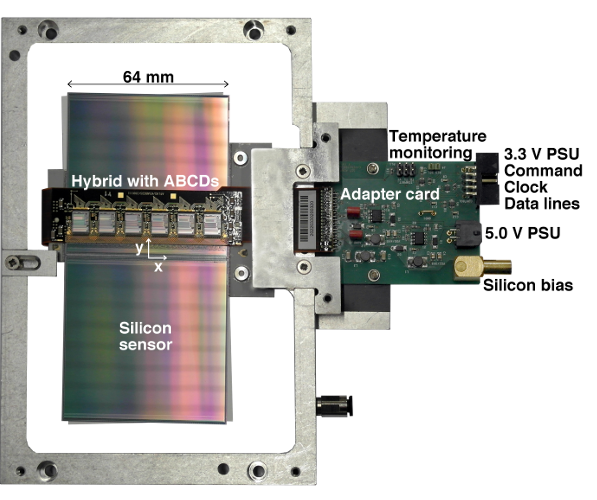}
  \caption{ATLAS SemiConductor Tracker module mounted on an aluminium handling frame with the adapter card.}
    \label{fig:sct}
\end{minipage}
\end{figure}

The module shown in Figure \ref{fig:sct} consists of two layers of single-sided p-in-n silicon sensors, each containing 768 strips with $80\,\mu$m pitch and mounted back-to-back on a thermal conductor. The layers are at a stereo angle of $\pm20\,$mrad, which gives position information in the $y$-direction. The active area of the sensors is $128\,\times\,64\,$mm. The  module is designed to have low mass and radiation length, which reduces muon scattering in the tracker to a minimum. The silicon strips are read out by the ABCD application specific integrated circuits \cite{abcd} mounted on a hybrid. These ABCDs implement signal amplification, analog to digital conversion, data compression and masking of noisy strips.  As the module is designed for the ATLAS experiment, the ABCDs are clocked at the CERN Large Hadron Collider (LHC) bunch crossing frequency of $40\,$MHz. Due to a power consumption of $6.0\,$W per module, cooling is required to prevent thermal runaway and to reduce thermal noise. Some cooling requirements could be avoided if the same silicon sensors were read out by custom electronics that were not tied to the relatively high LHC clock frequency. The unirradiated sensors are fully depleted at a $150\,$V reverse bias voltage.  \par

\subsection{Module adapter card}
\label{sec:adaptercard}

The adapter card has been designed to provide an interface between each silicon module and the readout board. The card is mounted on the aluminium frame of a module and connected to the hybrid using a 36-pin connector, as shown in Figure \ref{fig:sct}. Two switching power regulator circuits provide $3.5\,$V at $900\,$mA and $4.0\,$V at $400\,$mA from a $5\,$V input. The silicon bias voltage is connected to the adapter card via two LEMO connectors, which allows the bias voltage to be daisy chained between the modules. The silicon bias voltage is passed through an LC filter on the card before reaching the module, which contains a further double RC filter \cite{barrel}. LVDS line drivers and receivers on the card ensure the signal integrity between the readout board and silicon module, and help to reduce the risk of electrical damage to the ABCDs through the external connections. Five twisted pair cables between the readout board and adapter card carry two LVDS serial data lines, an LVDS serial command line, a clock line and a $3.3\,$V line to power the buffers. Finally, two pairs of pins on the card connect the thermistors on the module hybrid to an external temperature interlock.\par

\subsection{Readout board}
\label{sec:digitalboard}

\begin{figure} [t!]
\centering
\begin{minipage}[b!]{.22\textwidth}
  \centering
  \includegraphics[width=1.\linewidth]{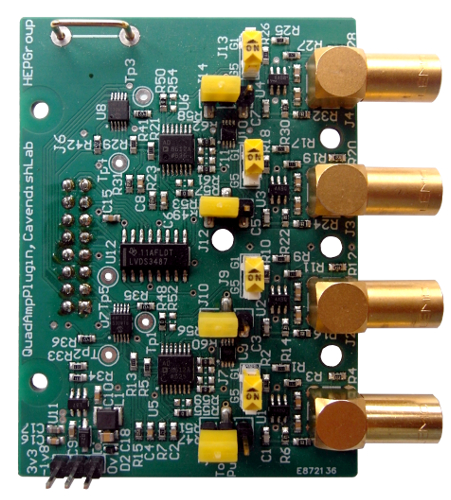}
  \caption{Hardware plugin for the trigger controller.}
  \label{fig:plugin}
\end{minipage}
\begin{minipage}{.05\textwidth}
\quad
\end{minipage}
\begin{minipage}[b!]{.68\textwidth}
  \centering
  \includegraphics[width=1.\linewidth]{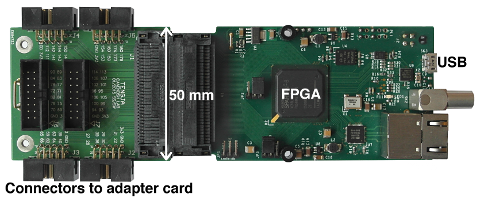}
  \caption{The readout board (right) and IO expander (left).}
    \label{fig:chimaera}
\end{minipage}
\end{figure}

The readout board shown in Figure \ref{fig:chimaera} is based on a Xilinx Spartan-6 XC6SLX25 field-programmable gate array (FPGA) \cite{XC6SLX25} with external USB and Ethernet connectivity. Originally designed by one of the authors to test compact arrays of single-photon-sensitive sensors for the LHCb Ring-Imaging CHerenkov (RICH) detector upgrade \cite{richtb}, the flexible design is adapted to the MST tracker by some modifications to the FPGA firmware and the addition of a passive IO expander.\par

The FPGA  provides ample logic, memory and IO resources to implement the required functions. The design occupies about $45\,$\% of the available logic slices and about $50\,$\% of the available block RAM. The wide range of configurable IO standards facilitates the interfacing to the front-end electronics. Ethernet connectivity is provided via the Micrel KSZ9021R transceiver \cite{KSZ9021R} and USB connectivity with the Cypress CY7C68013A peripheral microcontroller \cite{CYC68013A} and the Digilent Adept runtime environment \cite{Digilent}. \par

The FPGA firmware is designed for the readout to work autonomously once configured at the start of a run. Configuration data are sent as a single Ethernet TCP/IP packet with custom protocol and containing a unique target identifier in the packet payload. An Ethernet compliant Media Access Controller is not implemented in the FPGA. The firmware ignores any packets with a different protocol number or a unique identifier that does not match the hardware. Once validated, firmware in the FPGA converts the configuration data into a stream of serial commands to the ABCDs to set the operating parameters and to enable data acquisition mode. Serial data returned by the ABCDs are captured and buffered in FPGA RAM. The events are formatted into multi-event data packets and Ethernet and TCP/IP headers are added. The TCP/IP packets are then forwarded, employing fragmentation where needed and using the RGMII-compliant interface of the external transceiver. For these studies, the Ethernet interface is operated in 100-baseTX mode, but the hardware supports 1000-baseT mode with minor firmware changes. \par

A readout board with a different version of FPGA firmware is used to distribute the readout trigger. An additional hardware plugin, shown in Figure \ref{fig:plugin}, is connected to the IO expander and performs the discrimination and coincidence of the LEMO trigger signals from the scintillator detectors. In this configuration, which is referred to as the \textquoteleft trigger controller', the readout board is controlled through the USB interface rather than the Ethernet interface. The Cypress microcontroller provides the physical USB2 transceiver. It is programmed with the Digilent Adept framework that provides a USB to 8-bit asynchronous parallel port interface (DEPP) and USB to JTAG interface (DJTG). The DEPP interface is used for general configuration and monitoring and the DJTG interface is used to program the FPGA and its associated flash memory. The Digilent Adept Application Programming Interface, provided by the vendor, is integrated with a Java\texttrademark\,GUI using the Java\texttrademark\,Native Interface.

\subsection{Trigger and readout}
\label{sec:trigger}

\begin{figure}[b!]
\centering
\includegraphics[width=0.85\linewidth]{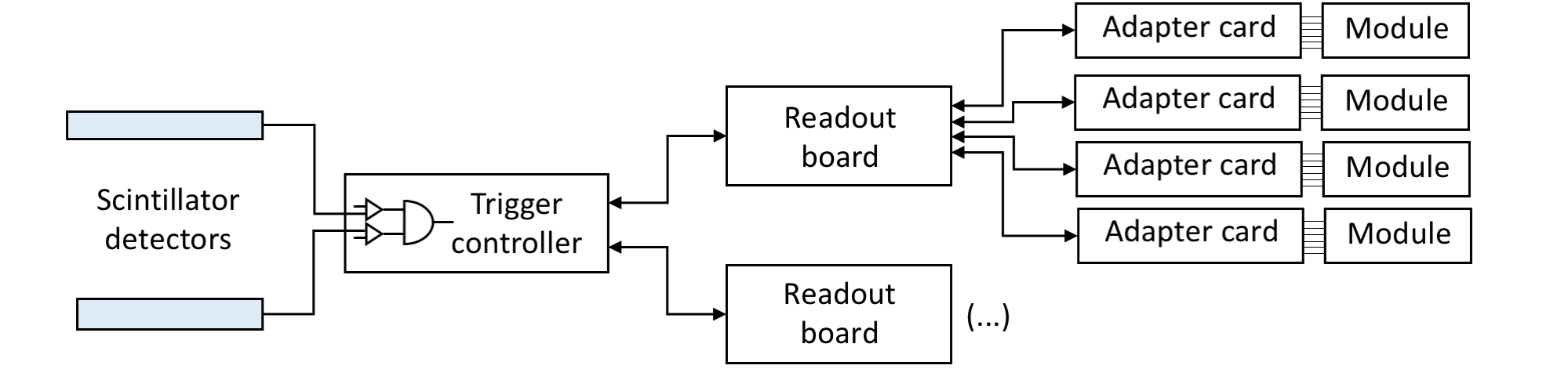}
\caption{\label{fig:trigger} Schematic of the electronic readout architecture. Two readout boards are used for the tracker, each connected to a station containing four modules.}
\end{figure}

\begin{figure}[b!]
\centering
\includegraphics[width=0.9\linewidth]{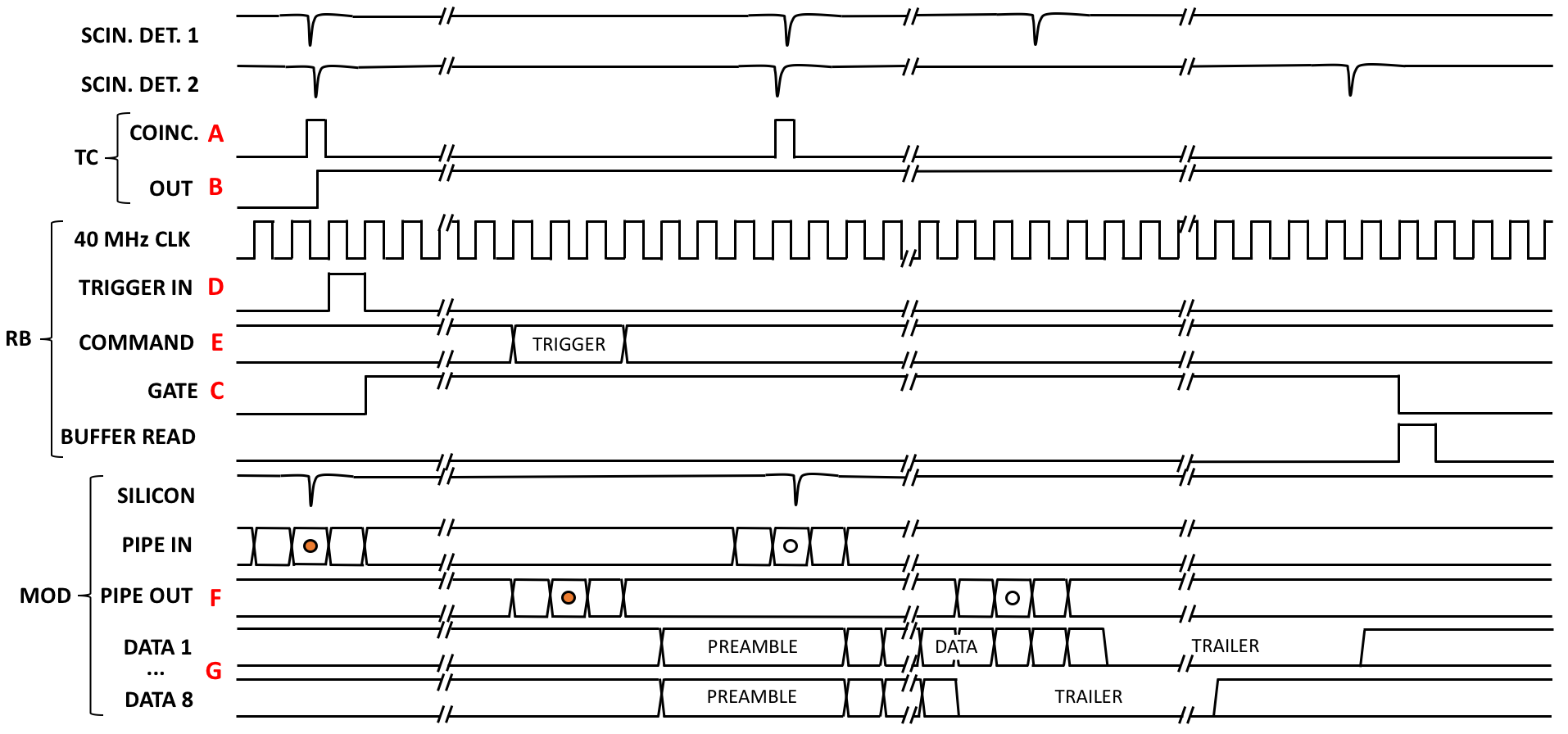}
\caption{\label{fig:signals} Timing diagram for signals in the trigger controller (TC), readout board (RB) and modules (MOD).}
\end{figure}

An overview of the electronic readout system is given in Figures \ref{fig:trigger} and \ref{fig:signals}. When the two scintillator detectors result in a coincidence (A in Figure \ref{fig:signals}), the trigger controller output to the readout board toggles (B). While the readout boards are processing, a gate is returned to the trigger controller (C), which prohibits any further triggers to be sent. The trigger controller output is registered by the readout board (D) on the next clock edge of the $40\,$MHz clock. On receipt of the trigger, the FPGA logic counts down a user-defined number of clock cycles before it issues a trigger command (E) to all ABCDs. This delay synchronises the trigger command with the arrival of the silicon signal at the end of the ABCD pipeline (F). The ABCDs read three consecutive $25\,$ns time bins at the end of the pipeline. Although the arrival of a cosmic ray muon is asynchronous to the $25\,$ns clock, the ability to stretch the signal to more than $25\,$ns using the ABCD shaping amplifier, combined with the three-bin readout, still results in a high detection efficiency. \par

The modules respond to the trigger command by sending out data (G) on the eight serial data lines connected to a readout board. The FPGA logic monitors the serial data lines for the preamble pattern following receipt of the trigger (D). When the pattern is recognised, the sequence of the preamble, data and trailer are written into a buffer in the FPGA RAM. The gate is deasserted and the readout of the buffer via Ethernet is started only after the trailer has been received on all serial data lines. \par

The readout buffers are chosen to be $16\,$kbit in size, which is used to buffer 32 events in $512\,$bit blocks. The buffering allows the bandwidth of the downstream electronics and Ethernet to be used efficiently, while preserving the ability to maintain a higher instantaneous trigger rate. As the ABCD data are zero suppressed, the length of the data stream depends on the occupancy in the silicon. The chosen buffer size guarantees that at least 28 hits per silicon layer can be stored per event. In the rare case of a buffer overwrite, the buffer address pointer wraps around and the event gets rejected in software, as the preamble pattern is no longer present. The Ethernet packets from the two readout boards are routed through an Ethernet switch to the control PC. When the CPU usage of the PC is high, Ethernet packages can occasionally get lost. Although a small rate of event loss is tolerable, the synchronisation between the tracker stations is lost. The event IDs in the Ethernet headers from both readout boards are therefore compared, and if these are not the same, packets are discarded until synchronisation is restored. \par 

The trigger rate is ultimately limited by the time taken by the ABCDs to transmit the hit information, which is done at one bit per $25\,$ns clock period. The shortest ABCD pattern is 53 bits at zero occupancy, giving a maximum trigger rate around $750\,$kHz. The longest ABCD pattern, at $50\,$\% occupancy with alternating hit strips, is 6563 bits, corresponding to a trigger rate of $6\,$kHz. For the chosen readout buffer size, the $100\,$MBit$\,$s$^{-1}$ Ethernet link limits the trigger rate to $195\,$kHz. In practise, the downstream performance of the Ethernet switch and control PC also have to be taken into account. The tracker was successfully operated at an event rate of $30\,$kHz when it was used as a beam telescope during CERN LHCb upgrade beam tests. During cosmic ray data taking, the scintillator detectors trigger the tracker at approximately $2.5\,$Hz, which is well below the limit.\par 

\subsection{Threshold scan}
\label{sec:threshold}
\begin{figure}[h!]
\centering
\includegraphics[scale=0.45]{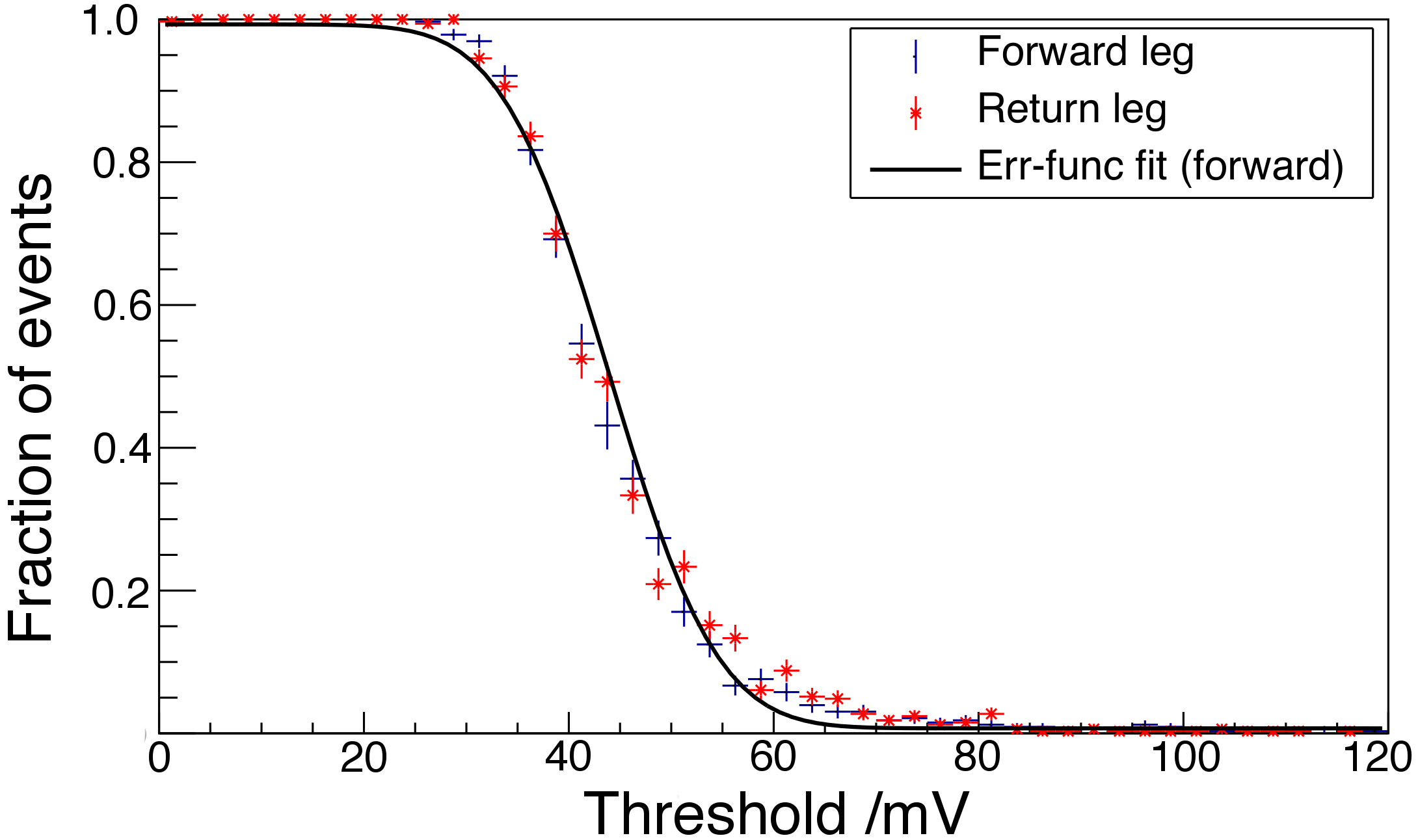}
\caption{Threshold scan to measure the silicon noise characteristics, showing no hysteresis for increasing and decreasing threshold steps.}
\label{fig:threshold}
\end{figure}

In order to measure the silicon noise characteristics, the threshold for analog to digital conversion in the ABCDs is incremented in steps of $2.5\,$mV and data are recorded at each step using a periodic trigger at $100\,$Hz frequency. A typical \textquotedblleft S-curve\textquotedblright\,response is shown in Figure \ref{fig:threshold}. A binomial fit of an error function is used to extract the mid point and width of the distribution, which for a typical silicon layer does not exceed $80\,$mV and $20\,$mV respectively. The average gain of the modules is given in \cite{barrel} as $55\,$mV$\,$fC$^{-1}$, which for a minimum ionising particle results in a signal of approximately $210\,$mV. However, whether the signal goes above threshold depends on the ABCD shaping amplifier configuration and the time of arrival of the cosmic ray signal with respect to the ABCD clock edges. For the studies in this paper, a global threshold of $137.5\,$mV applied to all modules results in a low noise environment. The efficiency is measured by fitting a track, as described in Section \ref{sec:track}, to all layers of silicon except the layer of interest and extrapolating this track to the layer of interest, to look for a hit in a window of 10 silicon strips around the predicted position. The obtained efficiency ranges from $86\,$\% to $93\,$\% and can be improved by optimising the threshold settings.\par

\clearpage

\subsection{Mechanics}
\label{sec:tracker}

\begin{figure}[htbp]
\centering 
\includegraphics[width=.9\textwidth]{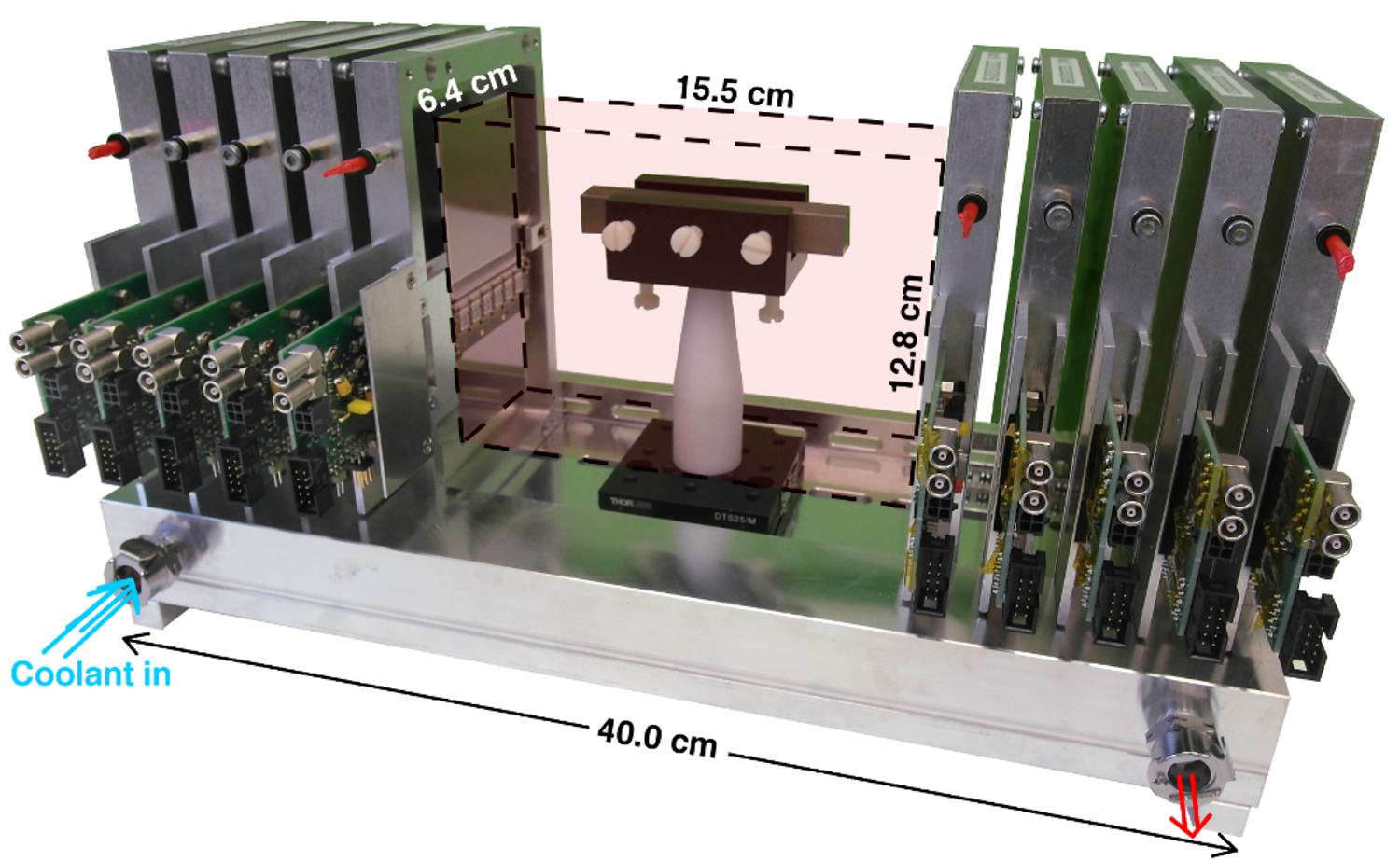} 
\caption{\label{fig:tracker} Aluminium support for cooling and alignment of the modules. To show the sensor, the aluminium cover plate has been removed from one of the modules. A sample is clamped in the plastic holder in the imaging region, which is marked in red. Each tracking station comprises four modules plus a spare.}
\end{figure}

The aluminium support in Figure \ref{fig:tracker} cools and aligns the modules. The spacing between the modules on the support can be varied, allowing the tracker to be optimised for high resolution at large spacing or greater muon acceptance at small spacing. The aluminium surfaces are precision manufactured to around $20\,\mu$m, which is a quarter of the silicon strip pitch, and constrain the modules to be aligned and parallel. Two tracking stations, each containing four modules plus a spare, are mounted on the support. In this study, the module spacing is $23.8\,$mm with a $155.2\,$mm separation between the tracking stations. The support can lie flat in a particle beam for example, or stand for maximal cosmic ray flux. \par

Water coolant at $10\,^{\circ}$C flows through the support at around 5-$10\,$L$\,$min$^{-1}$ to dissipate heat from the modules. The modules are orientated such that the corner with the shortest thermal path to the sensors is in contact with the cold reservoir. The flat surfaces provide a good thermal contact even without thermal paste. The ATLAS upgrade temperature interlock \cite{interlock} is used to switch off the tracker when the temperature at the module hybrid exceeds $41\,^{\circ}$C. This avoids potential damage to the sensors due to thermal runaway in case of failure of the cooling system. The tracker is operated in a light-tight box, which is flushed with nitrogen to avoid condensation on cold parts. Many other tracking geometries can be achieved with redesigned mechanics, where a sample can be placed between two tracking stations as long as these stations are well aligned. \par

The system is found to be stable over weeks of data taking, with temperature fluctuations within $0.3\,^{\circ}$C and with a consistent trigger rate, efficiency of the silicon sensors and scattering angle resolution. \par

\clearpage

\section{Performance of the tracker}
\label{sec:fitting}

Software has been developed to analyse the tracker output. As presented in Section \ref{sec:track}, adjacent strips above threshold are clustered and associated with muon tracks. The described algorithm is developed for higher multiplicity events when the tracker was operated in LHCb beam tests. Cosmic ray events typically contain one muon track and are therefore less challenging to identify. Data from $8000$ muons recorded in the absence of a sample are used to align the modules by a translation and rotation described in Section \ref{sec:softwareAlign}. A Geant4 simulation of the tracker is presented in Section \ref{sec:geant4} and is used to better understand the scattering angle resolution of the tracker, which is discussed in Section \ref{sec:res}. \par

\subsection{Track identification} 
\label{sec:track}

\begin{figure} [b!]
\centering
\begin{minipage}[t!]{.45\textwidth}
  \centering
  \includegraphics[width=1.\linewidth]{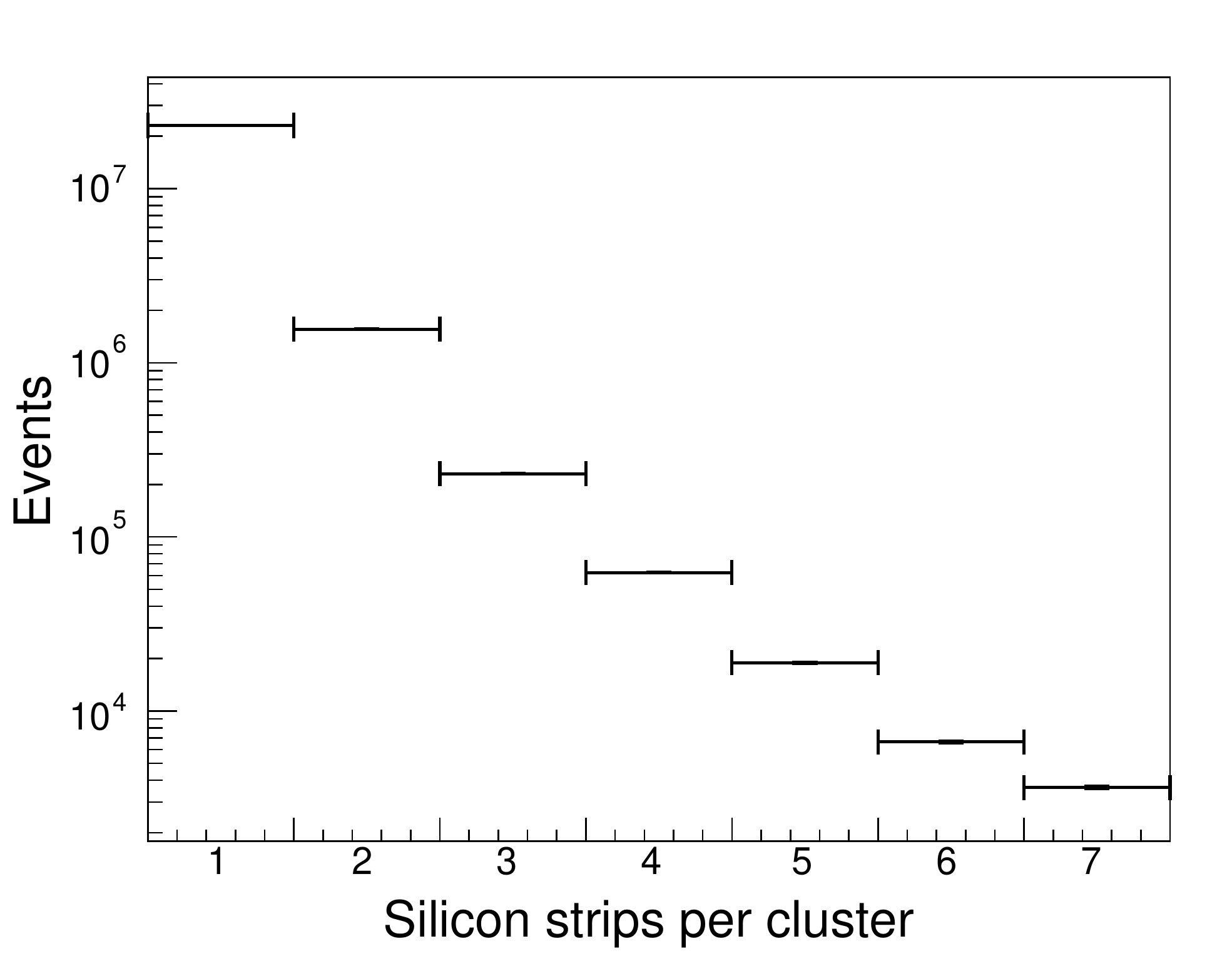}
  \caption{Number of adjacent silicon strips that make up a cluster.}
    \label{fig:cluster}
\end{minipage}
\begin{minipage}{.05\textwidth}
\quad
\end{minipage}
\begin{minipage}[t!]{.45\textwidth}
  \centering
  \includegraphics[width=1.\linewidth]{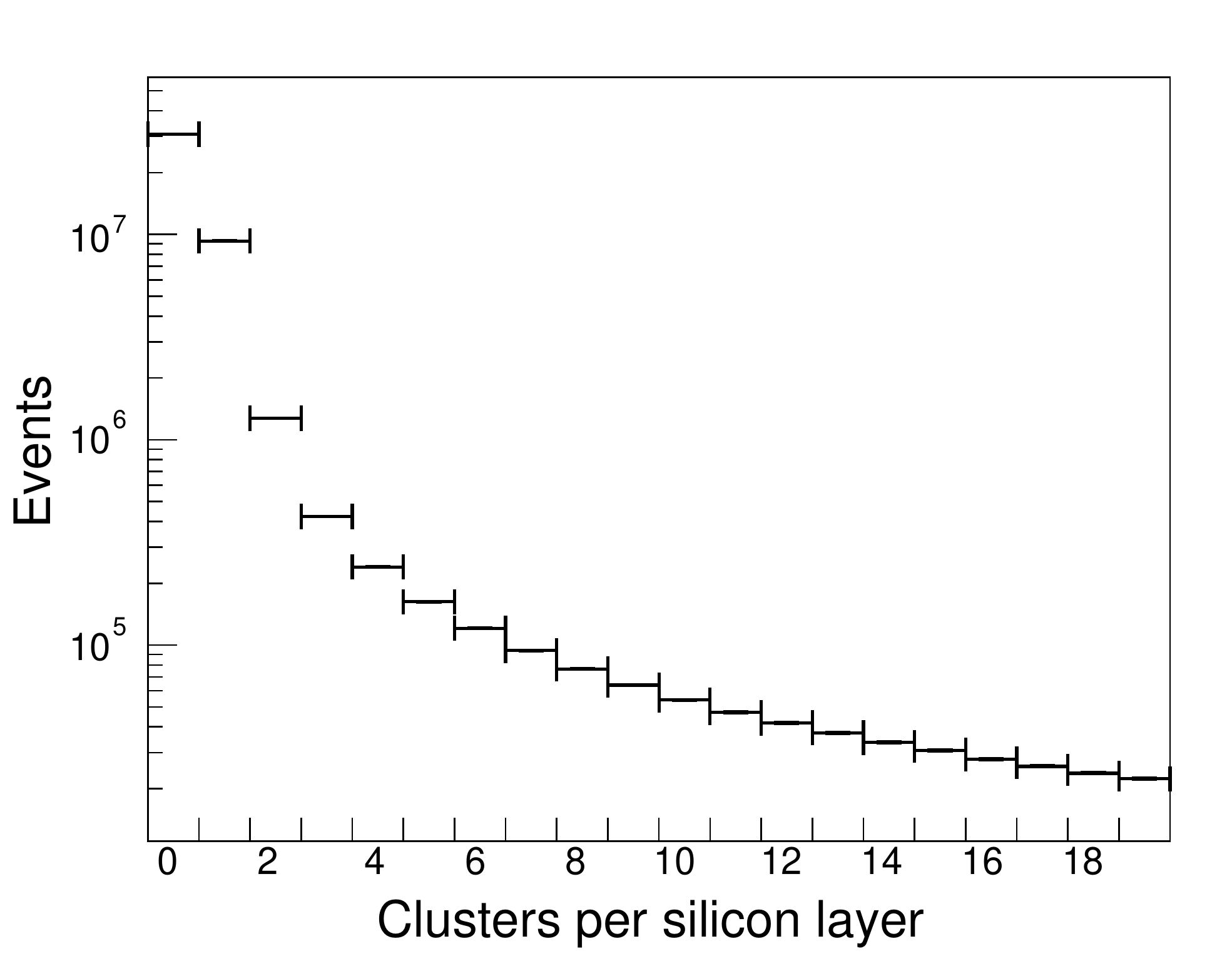}
  \caption{Number of clusters of hits per silicon layer in the tracker.}
  \label{fig:perlayer}
\end{minipage}
\end{figure}

A small number of silicon strips that produce a hit in more than $90\,$\% of the events are considered noisy strips and are therefore masked during data taking. The first step in the analysis is to cluster adjacent hit strips in the same silicon layer. The cluster position is assigned to the boundary between strips for an even number of hits or the centre of the central strip for an odd number of hits. The number of adjacent strips that form a cluster is shown in Figure \ref{fig:cluster}. The analysis discards $1.3\,$\% of the clusters on the basis that these contain more than two strips. The number of clusters per silicon layer in Figure \ref{fig:perlayer} is typically low, with a mean occupancy of $0.5$ clusters per layer per event in the absence of cosmic ray tracks.

\begin{figure}[t!]
\centering 
\includegraphics[width=0.7\textwidth]{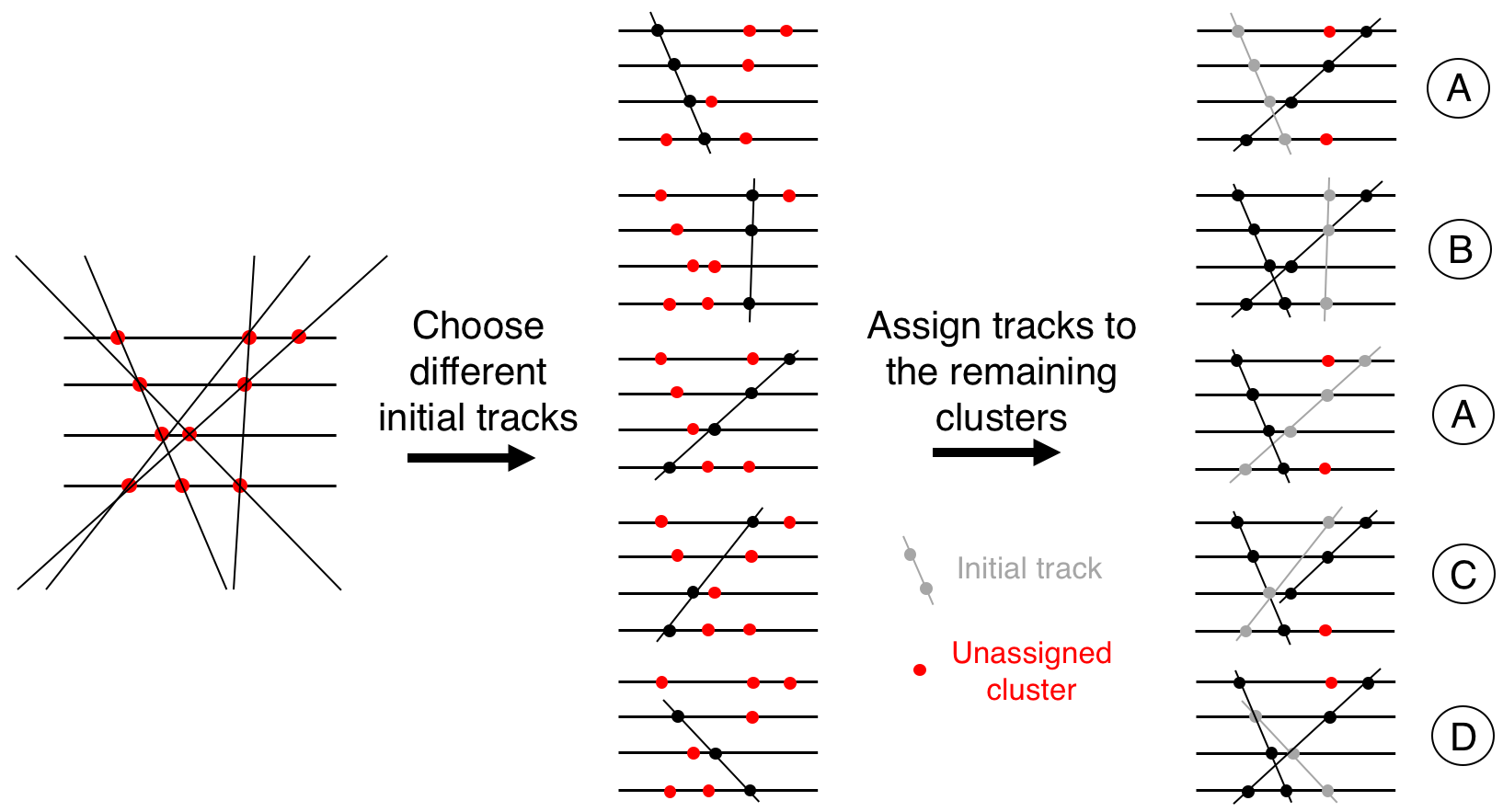} 
\caption{\label{fig:trackid} Example of track identification in software. Different initial tracks are chosen from the set of five potential tracks, and the associated clusters removed. Tracks are assigned to the remaining clusters, with preference to tracks with four clusters and tracks with lowest mean squared deviation. Configuration B contains three tracks and no unassigned clusters.}
\end{figure}

The even and odd silicon layers are treated separately in the fitting procedure due to the stereo angle. A pair of lines is therefore found for each tracking station using linear regression in the $xz$-plane, and combined using the stereo angle to obtain the gradient and intercept in the $yz$-plane. The first step in the association of clusters to tracks is to draw all possible tracks connecting three or four clusters in different modules within the acceptance, with a loose cut on the residuals to remove non-physical tracks. In the example in Figure \ref{fig:trackid}, this step is shown on the left hand side, where five tracks are created. However, some of these tracks share the same cluster, which is unlikely given the low muon track multiplicity and $80\,\mu$m silicon strip pitch. The algorithm therefore steps through the tracks and removes the associated clusters, as shown in the second column in Figure \ref{fig:trackid}. Each choice of initial track leaves a different set of remaining clusters. To assess which initial track gives the most likely set of tracks, additional tracks are assigned to the remaining clusters, with preference given to tracks containing four clusters followed by tracks with the smallest mean squared deviation. The associated clusters are removed, until it is no longer possible to draw tracks. \par

The most likely configuration of muon tracks is defined as the one that (i) leaves the minimum number of clusters unassigned. In configurations where the number of unassigned clusters is the same, preference is given to (ii) the one with the lowest mean squared deviation. It is also favoured to have (iii) one less track if this only increases the number of unassigned clusters by one. In the example of Figure \ref{fig:trackid}, the best configuration is B, based on criterion (i), as this configuration assigns all clusters to three muon tracks. Based on criterion (iii), configuration A is preferred over C and D. \par 

\subsection{Software alignment of modules}
\label{sec:softwareAlign}

\begin{figure} [t!]
\centering
\begin{minipage}[t!]{.45\textwidth}
  	\centering
	\includegraphics[width=1.0\textwidth]{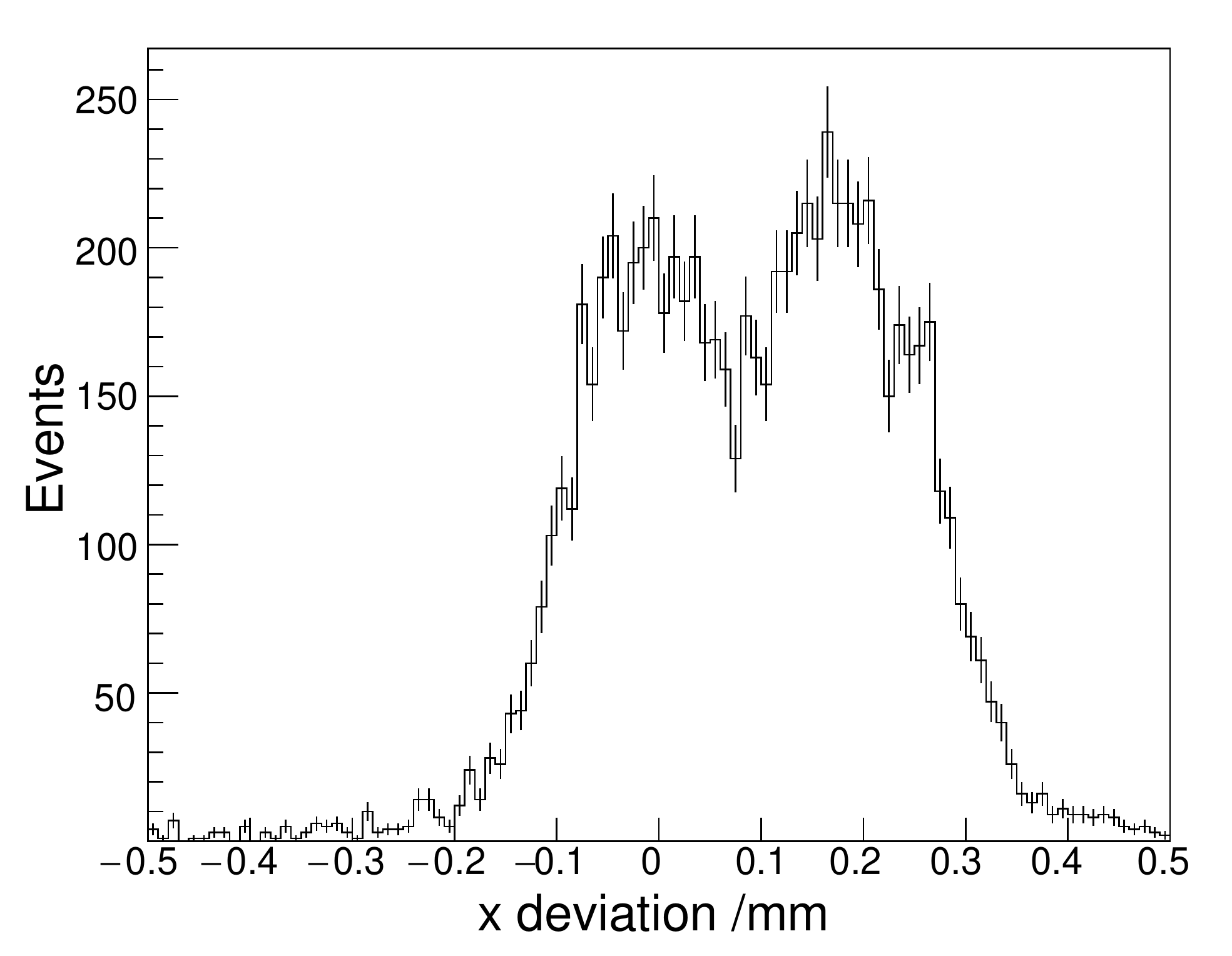}
	\caption{\label{fig:alignbefore} Deviation of clusters from the predicted positions in the $x$-direction, based on linear regression through all silicon layers excluding the layer of interest.}
\end{minipage}
\begin{minipage}{.05\textwidth}
	\quad
\end{minipage}
\begin{minipage}[t!]{.45\textwidth}
	\centering 
	\includegraphics[width=1.0\linewidth]{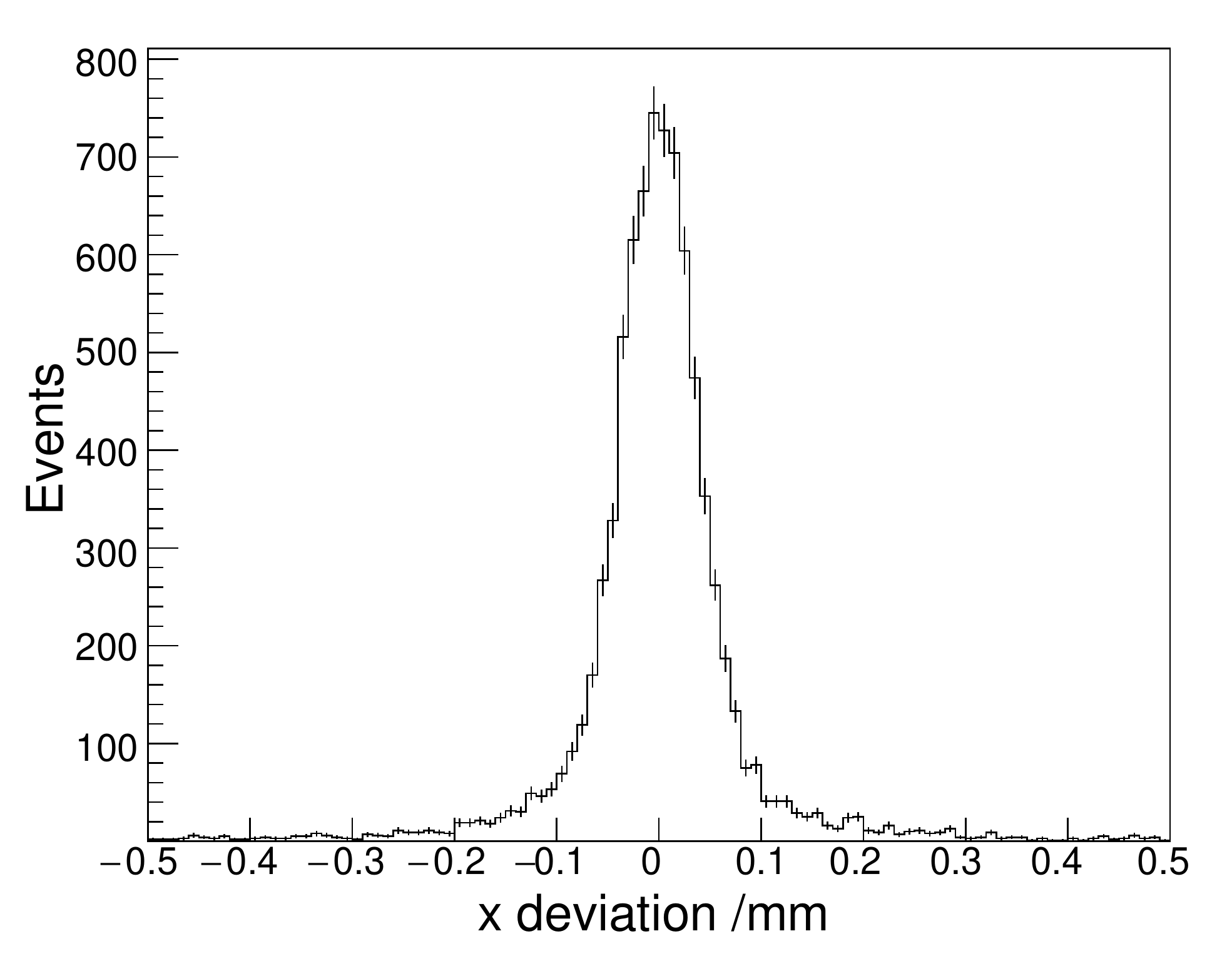}
  	\caption{\label{fig:alignafter} Identical plot to the one shown in Figure \ref{fig:alignbefore}, after correcting of the module alignment in software.}
\end{minipage}
\end{figure}

The alignment of the modules in the $x$-direction is corrected in software using data taken in the absence of a sample. Track identification is performed as described in Section \ref{sec:track}. However, when a muon track with a cluster in each silicon layer of both tracking stations is identified, the predicted position of the cluster on each layer is calculated from a linear fit through the fifteen other layers. The deviation from this predicted position on a typical layer is plotted for $8000$ muons in Figure \ref{fig:alignbefore}, and shows a double-peak structure with a non-zero mean. The modules are shifted by half of the deviation from the predicted position in the $x$-direction, which is iterated $20$ times, at which point the mean deviation from zero is typically $0.1\,\mu$m. After alignment, the angle between the line connecting the centres of the layers and the normal to the layers is less than $0.5\,$mrad and no further correction is applied. A rotation of modules around the centre in the $xy$-plane results in deviations in the $x$-direction that are different for positive and negative $y$-coordinates of the clusters. This creates the two-peak structure in Figure \ref{fig:alignbefore}, which is corrected by the same iterative approach as the translation. \par

The deviation from the predicted position after alignment is shown in Figure \ref{fig:alignafter}. The standard deviation of the distribution is $43\,\mu$m due to internal scattering of the relatively low energy cosmic ray muons used for the alignment. Typically, the correction to the mean in the $x$-direction was below $100\,\mu$m. Because a similar misalignment in the $x$ and $y$-direction is expected from the mechanics, it is decided not to correct the alignment in the $y$-direction, as the spatial resolution in this direction is around $2\,$mm. The moderate angular acceptance of the tracker limits the error in the $x$-direction due to misalignment in the $z$-direction, which is therefore also not corrected in the analysis.
\par

\subsection{Geant4 detector description} 
\label{sec:geant4}

The angular resolution of the tracker is limited by internal scattering, which depends on the muon momentum, the angle of incidence, the material budget of the module and the number of modules in the tracker. Additionally, the strip width, any remaining misalignment after calibration, inefficiencies of the modules and cross talk or electronic noise reduce the fit 	quality of the tracks. To better understand some of these contributions, a description of the tracker is created in Geant4. 

The tracker geometry is simulated by modules in air with a random offset in the $x$-direction sampled from a Gaussian with standard deviation of $80\,\mu$m and a rotation up to $3\,$mrad in the $xy$-plane. The sensors are designed to have a low radiation length, and contribute $1.17\,$\%$\,$X$_{0}$ averaged over each sensor \cite{barrel}. Additionally, the protective covers on either side of the silicon introduce $1.0\,$mm of aluminium with a radiation length of $1.12\,$\%$\,$X$_{0}$. To simulate the spatial resolution of the sensors, each hit in the sensor is assigned to the centre of an $80\,\mu$m strip. The output from the strip assignment has the same format as the experimental data file, which means that identical track identification, calibration and image reconstruction can be used in both cases. \par

The cosmic ray muon source is modelled as a planar source, centred with the tracker and positioned $2.9\,$mm above the top silicon sensor. The source has an area of $66.6\times132.6\,$mm$^2$, as muons generated outside this area cannot go through the tracker. For $100\,$\% efficient silicon sensors, $3.7\,$\% of the source particles result in a track through the tracker due to the acceptance. At $90\,$\% silicon efficiency, $2.6\,$\% of the source particles result in a reconstructed track. The cosmic ray angular and energy distributions from \cite{cosmicDistribution} are implemented.

\subsection{Scattering angle resolution results}
\label{sec:res}

\begin{figure} [b!]
\centering
\begin{minipage}[t!]{.45\textwidth}
  	\centering
	\includegraphics[width=1.0\textwidth]{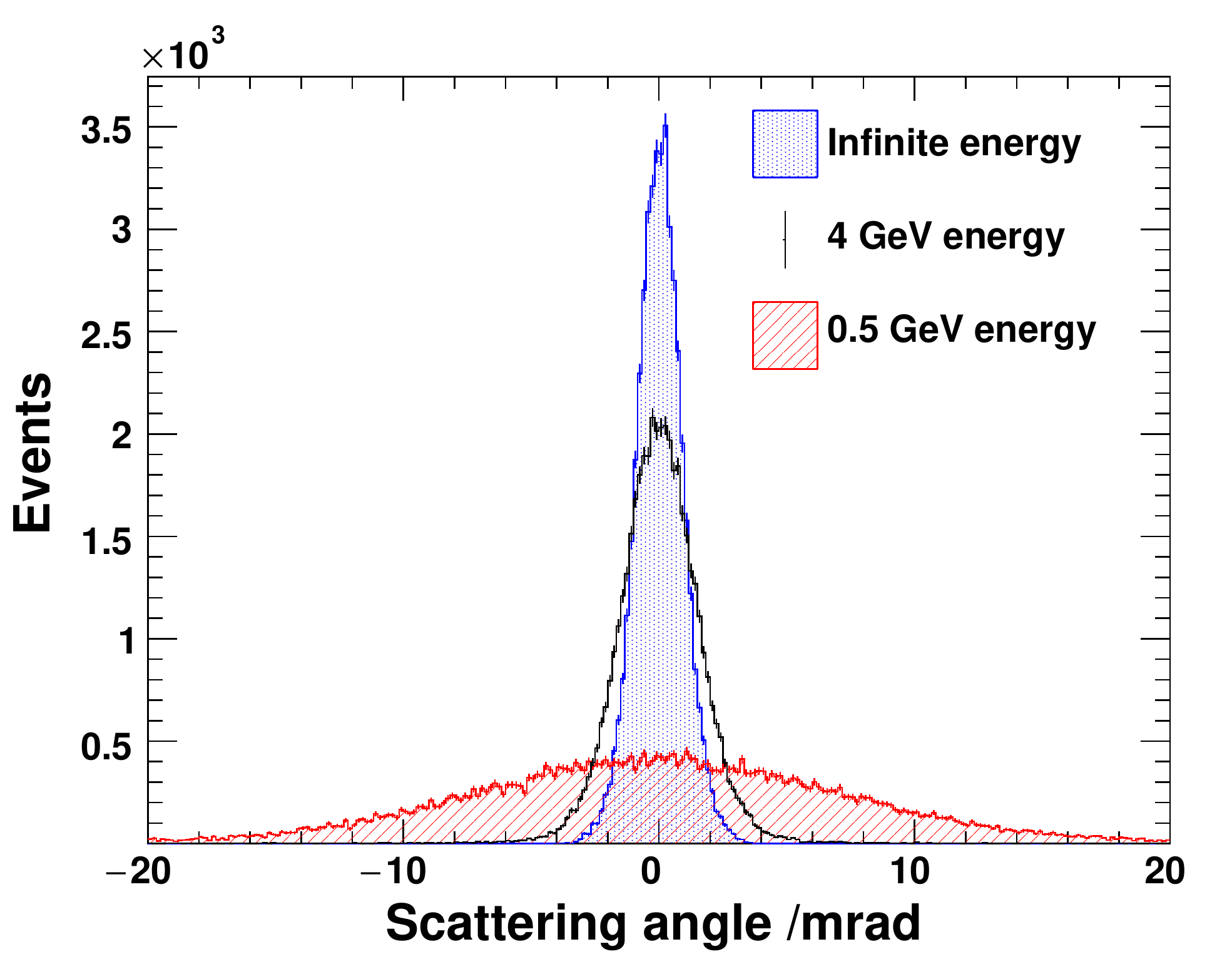}
	\caption{\label{fig:momentum} Simulated angle between incoming and outgoing muon tracks in the absence of a sample, for infinite, $4\,$GeV and $0.5\,$GeV energy.}
\end{minipage}
\begin{minipage}{.05\textwidth}
	\quad
\end{minipage}
\begin{minipage}[t!]{.45\textwidth}
	\centering 
	\includegraphics[width=1.0\linewidth]{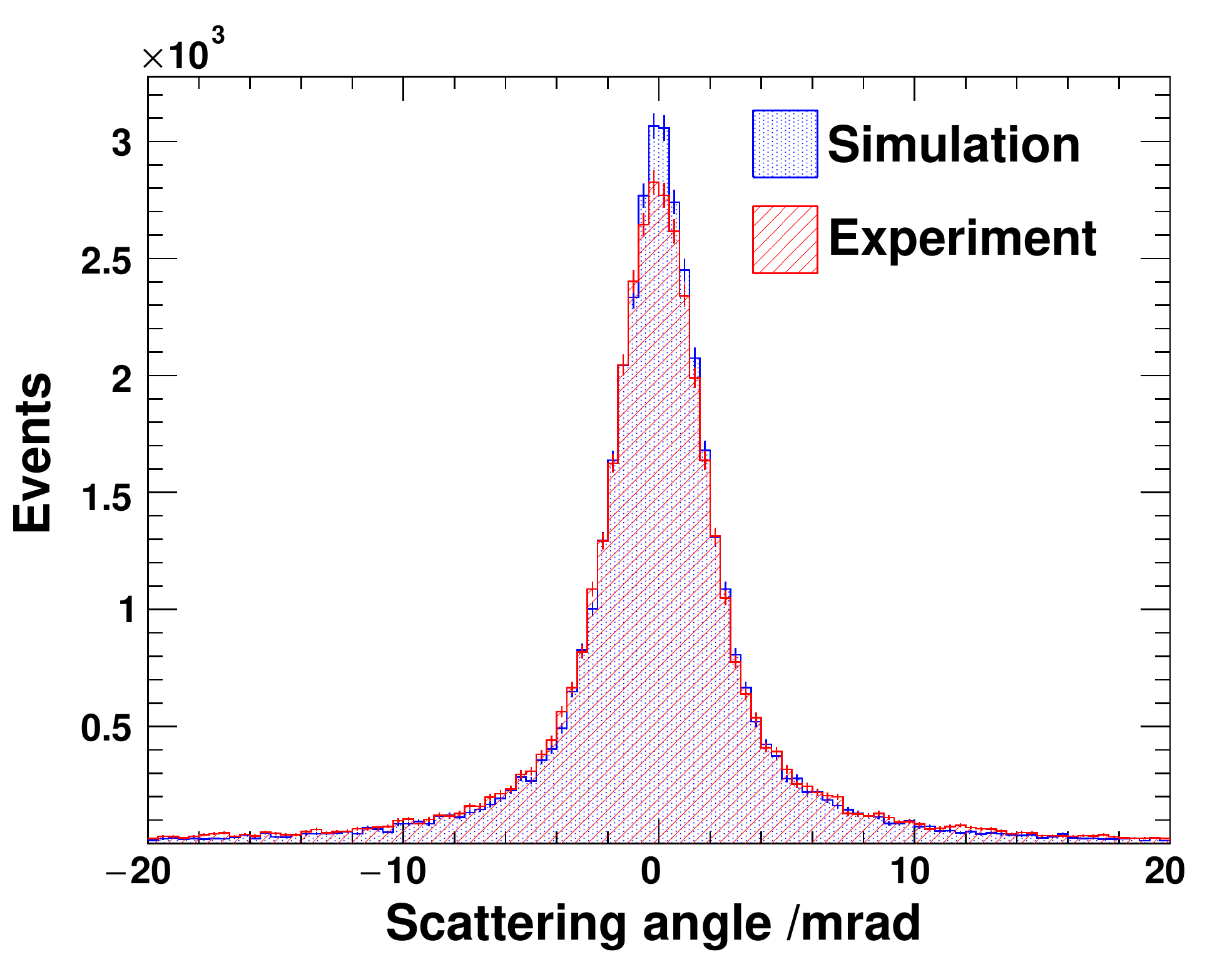}
  	\caption{\label{fig:simexp} Angle between incoming and outgoing muon tracks for 40000 scattering events in simulation and in experiment.}
\end{minipage}
\end{figure}

The tracker simulation is run for mono-energetic muons at the cosmic ray angular distribution, with infinite energy, with $4\,$GeV average cosmic ray energy and with $0.5\,$GeV energy. The angle between the incoming and outgoing muon tracks is shown in Figure \ref{fig:momentum}. The width of the Gaussian distribution for infinite energy is $0.9\,$mrad and arises from the silicon strip width and module alignment errors. This width increases to $1.5\,$mrad for $4\,$GeV muons and $7.3\,$mrad for $0.5\,$GeV muons due to internal scattering. \par

To compare the scattering angle distribution between simulation and experiment in Figure \ref{fig:simexp}, the simulation is run with the cosmic ray energy distribution and $40000$ events in the absence of a sample. The scattering angle distribution is a convolution of a Gaussian distribution, which has a width dependent on the cosmic ray energy as shown in Equation \ref{eq:scattering}, and the broad energy spectrum of cosmic ray muons. The benefit of reducing pixel dimensions and alignment errors is only significant for the narrowest Gaussians associated with the highest energy muons. \par

\begin{figure}[t!]
\centering 
\includegraphics[width=0.48\textwidth]{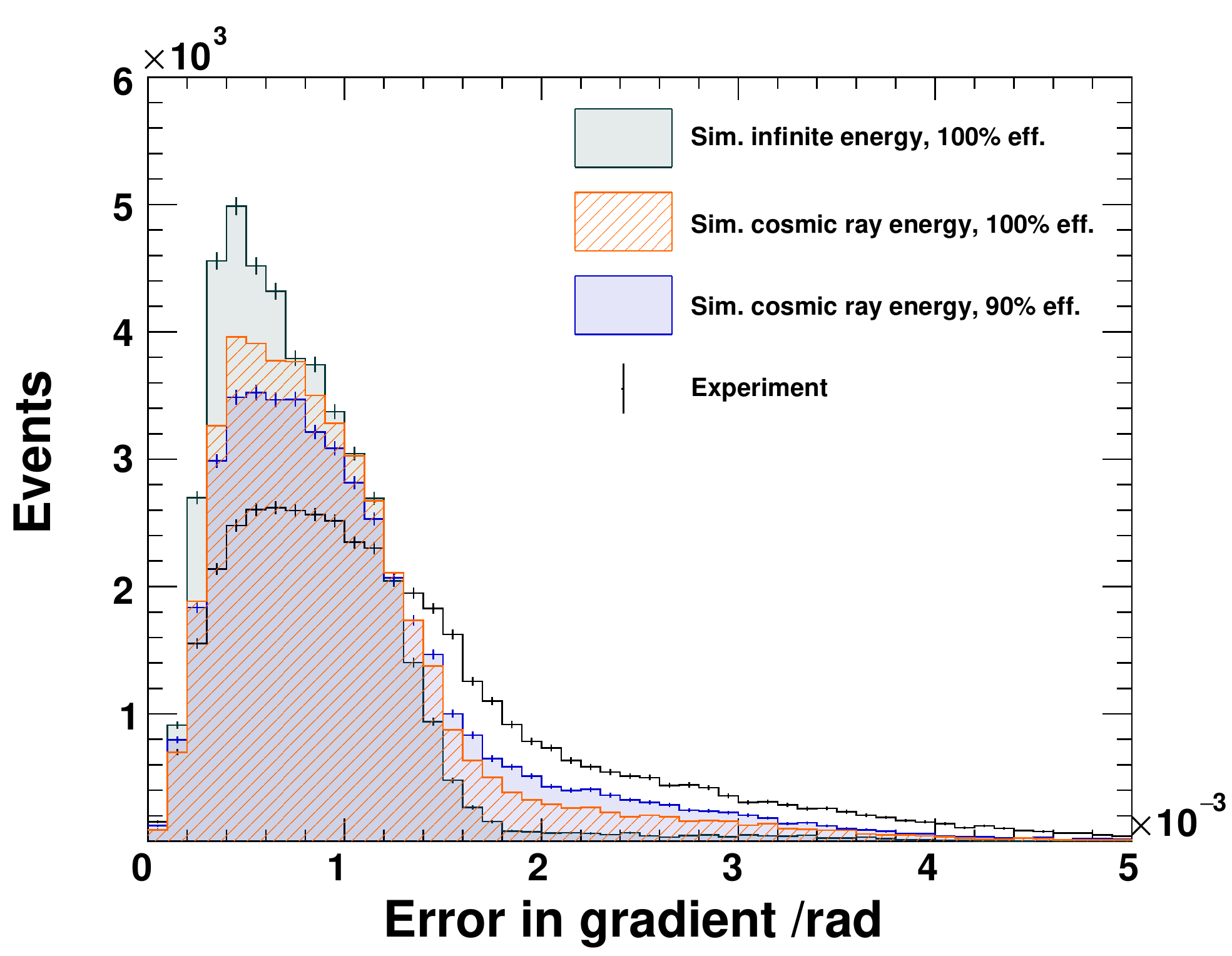} 
\caption{\label{fig:error} Error in the gradient of the straight line fit to the hit silicon strips in the tracker. The curves show experiment, simulation with infinite energy and simulation with the cosmic ray muon energy spectrum at two different silicon sensor efficiencies.}
\end{figure}

The error in the scattering angle is estimated using the error in the gradient of the straight line fit to hits in the silicon sensors, which is calculated following the approach in \cite{practical}. The results shown in Figure \ref{fig:error} show a mode of less than $1\,$mrad. The error from simulation with mono-energetic particles with infinite energy and $100\,$\% efficient silicon sensors shows the lower limit in the error from the strip width and small deviations in the module position. As expected, the error distribution broadens when the simulation is run with the cosmic ray energy distribution, and broadens further when inefficiencies are introduced in the silicon sensors. \par 

The experimental results show a slightly wider distribution in Figure \ref{fig:simexp} and a larger error in the gradient in Figure \ref{fig:error} than simulation. These results can be attributed to misalignments that are not corrected in the tracker calibration or to uncertainty in cluster assignment due to detector noise or cross talk for example.\par

\section{Image reconstruction}
\label{sec:results}

\subsection{Angle Statistics Reconstruction}
\label{sec:asr}

The Angle Statistics Reconstruction (ASR) algorithm is used for the image reconstruction in this study. ASR is chosen as it is robust to changes in the system setup, easy to implement and has been shown to perform better than Point of Closest Approach (PoCA) algorithms \cite{asr}. For a given volume element, the algorithm stores a score for each event that has the shortest distance between the volume element and the incoming and outgoing muon tracks below a threshold distance. After all data are processed, the events in the volume element are listed in ascending order by their score. The significance of each event in the list is weighted by the path length of the track through the volume element. Furthermore, to smooth the image, events in nearby volume elements are added to the list with a low weight of $30\,$\% for adjacent and $15\,$\% for diagonal volume elements. The event at a fixed fraction of the total weighted number of events is chosen, and the score stored in that event is used as the final score to create the ASR image. The fixed fraction is referred to as the \textquoteleft quantile'. For a detailed description of the algorithm, its motivation and performance see \cite{asr}.  \par

\begin{figure}[h!]
\centering 
\includegraphics[width=0.8\textwidth]{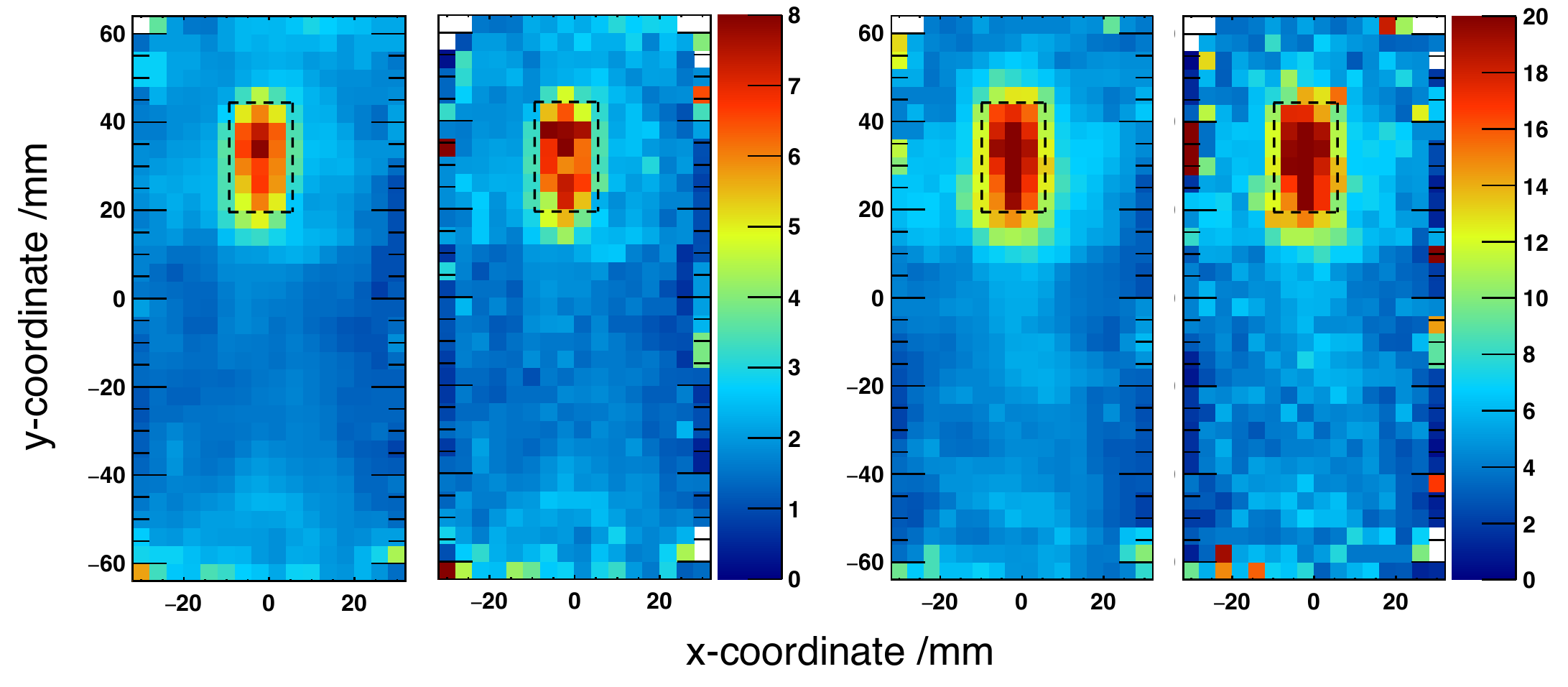} 
\caption{\label{fig:leadsmooth} ASR reconstruction of a lead sample with $45000$ muons from experiment. The black dotted lines show the sample contours. From left to right: smoothed and non-smoothed ASR, for a 0.5 quantile and for a 0.8 quantile.}
\end{figure}

The score for ASR in this study is the absolute value of the scattering angle in the $xz$-plane. The $2\,$mm spatial resolution in the $yz$-plane means that the scattering angle in this direction is not measured accurately enough for tomography, but the information is still used to localise the scattering event. Figure \ref{fig:leadsmooth} shows the effect of smoothing and quantile choice in the $xy$-projection of a $15\times25\times55\,$mm$^{3}$ lead sample imaged using $45000$ muons. As discussed in \cite{asr}, the algorithm performs well over a large range of quantiles, and in this paper a 0.8 quantile and $4\times4\times4\,$mm$^3$ volume elements are found to give best contrast between the sample and background.

\subsection{Cosmic ray imaging results}
\label{sec:iron}

\begin{figure}[h!]
\centering 
\includegraphics[width=0.8\textwidth]{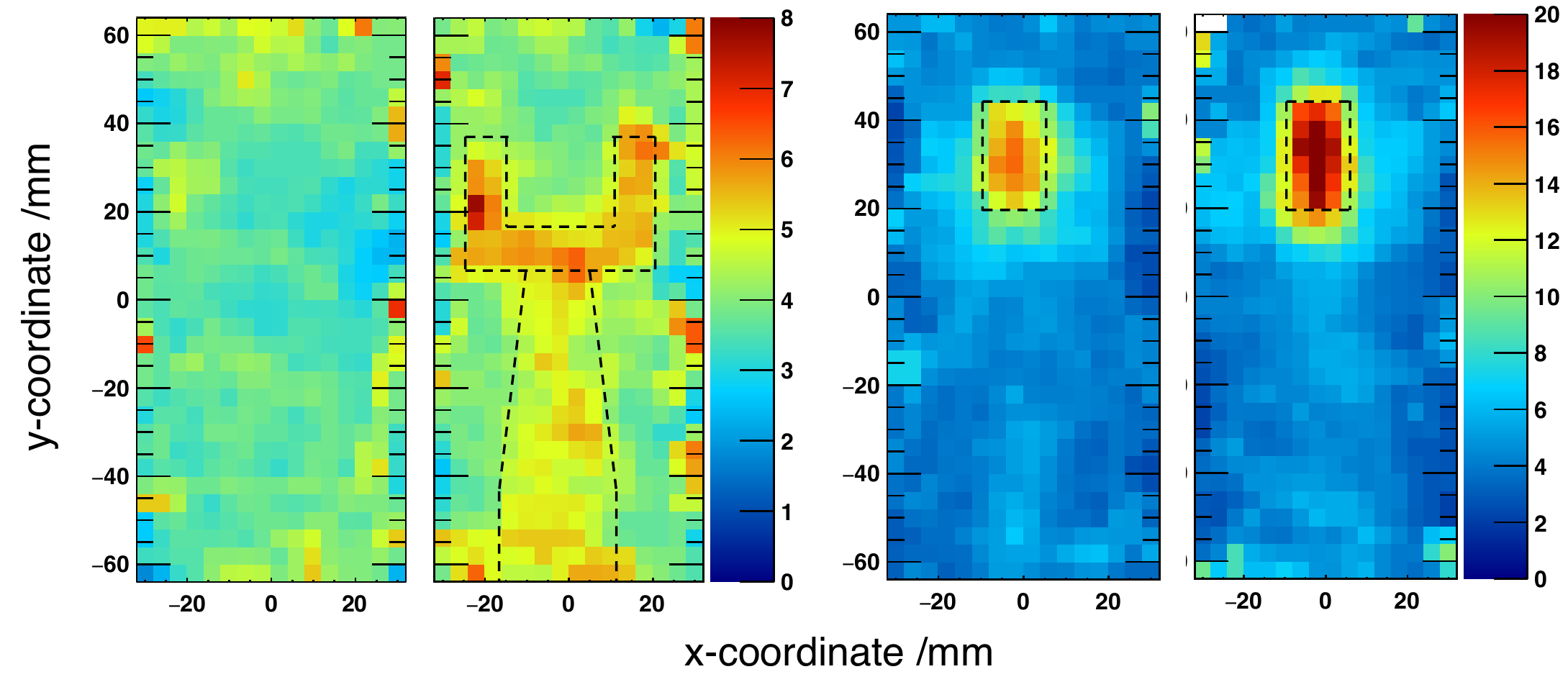} 
\caption{\label{fig:sample} ASR images with $45000$ muons from experiment. The black dotted lines show the sample contours. From left to right: air, plastic holder, iron and lead samples.}
\end{figure}

The $15\times25\times50\,$mm$^{3}$ iron and lead samples are clamped in the holder shown in Figure \ref{fig:tracker}. The holder is plastic in order to minimise background scattering and is mounted on a translation stage, which is outside the acceptance. The results in Figure \ref{fig:sample} demonstrate the ability to distinguish between air and the plastic sample holder. When introducing iron and lead samples, the multiple scattering significantly increases and a new colour scale is adapted. On this scale, the lead shows better visual contrast, as expected from its higher atomic number and density than iron. \par

\begin{figure}[h!]
\centering 
\includegraphics[width=0.6\textwidth]{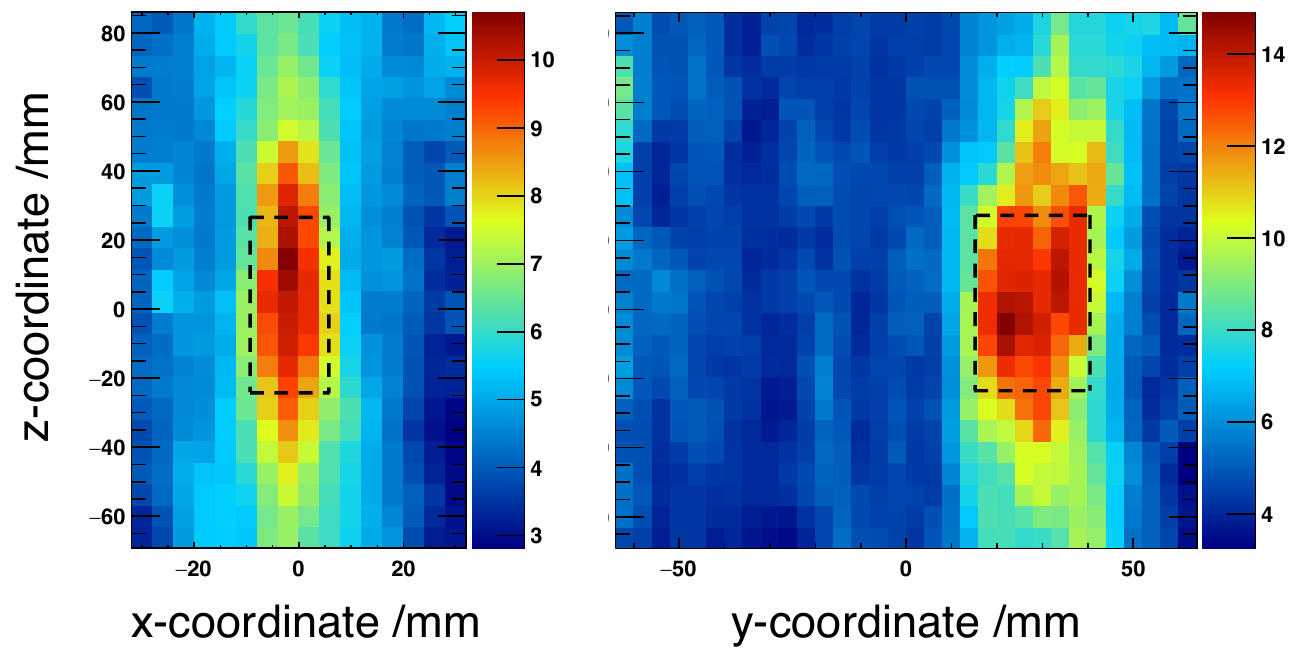} 
\caption{\label{fig:z} ASR images of a lead sample with $45000$ muons from experiment, projected captionin the $xz$-plane and the $yz$-plane. The colour scales are chosen to present the best contrast.}
\end{figure}

Projections of the lead sample in the $xz$-plane and $yz$-plane are shown in Figure \ref{fig:z}. As a result of the cosmic ray angular spectrum and the acceptance of the tracker, most muons tracks travel vertically through the tracker, resulting in the observed reduced definition in the $z$-direction. This effect is asymmetric in the $yz$-projection, where the side closest to the centre of the sampled volume gives a better representation of the sample size, which is superimposed as a black dotted line. Non-vertical tracks that pass the sample on the side away from the centre are more likely to go out of acceptance, resulting in the conical reconstructed image of the sample.\par

\subsection{Image recording time and tracker improvements}
\label{sec:efficiency}

The scanning time is a limiting factor of MST with cosmic ray muons, which have an approximate flux of $1\,$cm$^{-2}\,$min$^{-1}$ at sea level. The rate from the simulated source of area $88.3\,$cm$^2$ in \linebreak Section \ref{sec:geant4} is therefore approximately $1.47\,$s$^{-1}$. For $90\,$\% efficient silicon sensors, this results in an estimated track rate of $3.89\times10^{-2}\,$s$^{-1}$, which is an upper limit as it assumes a $100\,$\% efficiency of the scintillator trigger detectors. In experiment, $45841\pm241$ tracks were recorded over a time of $340.4\pm0.5\,$hours, which gives a track rate of $(3.74\pm0.02)\times10^{-2}\,$s$^{-1}$. \par 

\begin{figure}[h!]
\centering 
\includegraphics[width=0.9\textwidth]{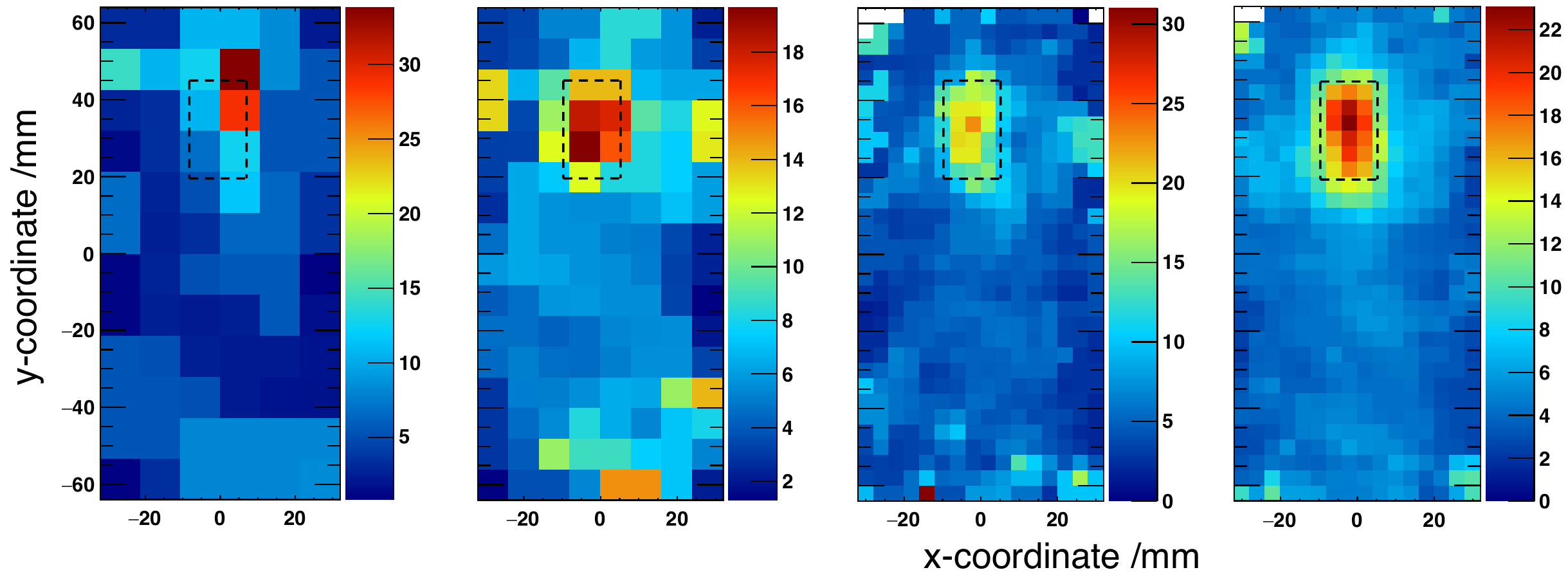} 
\caption{\label{fig:timeline} ASR images of a lead sample. From left to right: 100, 1000, 5000 and 25000 muons from experiment. The volume element size is increased at low statistics, in order to keep a sufficient number of muons in each element. The colour scales are chosen to present the best contrast.}
\end{figure}

Cosmic ray muon imaging has found many applications, some that can sacrifice high resolution imaging for short acquisition and object detection times \cite{rpc, drift}, and some that require high resolution images with less emphasis on the acquisition time \cite{nuclearwaste, pyramid}. An example of early detection of a lead sample is given in Figure \ref{fig:timeline} for 100, $1000$, $5000$ and $25000$ muons. The larger volume elements for 100 muons give an average of $1.4\,$muons per volume element. The relatively high score in the volume element at $x=-25\,$mm and $y=45\,$mm is therefore the effect of a scattering outlier. The lead sample stands out at less than $5000$ muons, which is likely to be reduced by improvements in image reconstruction and recognition algorithms. Heavier objects such as uranium are also expected to give higher contrast.\par

The rate from the tracker could be improved by optimising the threshold setting to achieve a silicon sensor efficiency close to $100\,$\%, which increases the track rate to $5.44\times10^{-2}\,$s$^{-1}$ estimated from simulation. Additionally, an array of sensors could be used, or the module position and the number of modules optimised, to increase the angular acceptance of the tracker. Although an array of sensors would require new mechanics and cooling, the readout electronics are designed to be able to scale to such a larger system. Finally, the amount of information per muon event would improve with a measurement of the muon momentum, or with a more accurate measurement of the scattering angle in the $yz$-plane, for example using pixel rather than strip sensors or moving to a crossed module geometry. \par

\clearpage

\clearpage

\section{Conclusion and outlook}
\label{sec:conclusion}

A compact, high resolution silicon strip tracker has been developed for cosmic rays muon scattering tomography. The combination of ATLAS SCT modules and LHCb RICH readout boards provides a scalable and potentially portable apparatus, which could therefore be extended to image larger objects or carried around to image smaller objects in situ. After track identification and software alignment of the muons, the muon scattering angle distribution obtained in experiment agrees with a Geant4 simulation of the detector. The simulation gives a scattering angle resolution of $0.9\,$mrad for high momentum muons and $1.5\,$mrad at the $4\,$GeV average energy of cosmic rays, which is limited by internal muon scattering. Images have been recorded using $45000$ muons and Angle Statistics Reconstruction, and show good contrast between air, plastic, iron and lead materials. A lead sample can be visually localised with less than $5000$ muons. Reconstruction and image processing algorithms are a key area for improvement in image quality \cite{wang1, wang2}, and the solid state tracker and corresponding Geant4 simulation provide a test-bed for further studies of reconstruction algorithms. It is also foreseen to find the optimal module positioning that maximises sensitivity in the presence of internal scattering, which can then be used for feature and material discrimination studies.

\acknowledgments
This work is funded, in part, by AWE, Aldermaston, Reading RG7 4PR, UK. AWE also loaned the scintillator detectors for the electronic readout trigger. The authors would like to thank Saevar Sigurdsson and the Cavendish Laboratory mechanical workshop for technical support.

\clearpage

\bibliographystyle{unsrt}
\bibliography{refs}
~\\~
\noindent{\textit{This document is of United Kingdom origin and contains proprietary information which is the property of the Secretary of State for Defence. It is furnished in confidence and may not be copied, used or disclosed in whole or in part without prior written consent of Defence Intellectual Property Rights DGDCDIPR-PL-Ministry of Defence, Abbey Wood, Bristol, BS34 8JH, England.}}

\end{document}